\documentclass[twocolumn,showpacs,showkeys,preprintnumbers,amsmath,amssymb,prc,superscriptaddress]{revtex4-1}

\usepackage{sistyle}
\usepackage[version=3]{mhchem}
\usepackage{xspace}
\usepackage{graphicx}
\usepackage{multirow}
\usepackage{setspace}
\usepackage{booktabs}		

\newcommand{\tref}[1]{Table~\ref{#1}}
\newcommand{\fref}[1]{Fig.~\ref{#1}}
\newcommand{\sref}[1]{Sec.~\ref{#1}}

\newcommand\Ar{\ce{^{31}Ar}\xspace}
\renewcommand{\S}{\ce{^{30}S}\xspace}
\newcommand\Cl{\ce{^{31}Cl}\xspace}
\renewcommand\P{\ce{^{29}P}\xspace}

\newcommand\SIE[3]{\num{#1}(\num{#2})\,\SI{}{#3}}
\newcommand\SIEb[3]{\num{#1}(\num{#2})$-$\SI{}{#3}}
\newcommand\SIb[2]{\num{#1}$-$\SI{}{#2}}
\newcommand\GG{$\Gamma_p/\Gamma_{\gamma}$\xspace}
\newcommand\GGr{$\Gamma_{\gamma}/\Gamma_p$\xspace}
\newcommand\rec{$\ce{^{29}P}(p,\gamma)\ce{^{30}S}$\xspace}

\begin{document}
\title{Relative proton and $\gamma$ widths of astrophysically important states in \S studied in the $\beta$-decay of \Ar.}

\author{G. T. Koldste}
\affiliation{Department of Physics and Astronomy, Aarhus University, DK-8000 Aarhus C, Denmark}
\author{B. Blank}
\affiliation{Centre d'{\'E}tudes Nucl{\'e}aire de Bordeaux-Gradignan, CNRS/IN2P3 -- Universit{\'e} Bordeaux I, F-33175 Gradignan Cedex, France}
\author{M. J. G. Borge}
\affiliation{Instituto de Estructura de la Materia, CSIC, E-28006 Madrid, Spain}
\author{J. A. Briz}
\affiliation{Instituto de Estructura de la Materia, CSIC, E-28006 Madrid, Spain}
\author{M. \surname{Carmona-Gallardo}}
\affiliation{Instituto de Estructura de la Materia, CSIC, E-28006 Madrid, Spain}
\author{L. M. Fraile} 
\affiliation{Grupo de F{\'i}sica Nuclear, Universidad Complutense, E-28040 Madrid, Spain}
\author{H. O. U. Fynbo}
\affiliation{Department of Physics and Astronomy, Aarhus University, DK-8000 Aarhus C, Denmark}
\author{J. Giovinazzo}
\affiliation{Centre d'{\'E}tudes Nucl{\'e}aire de Bordeaux-Gradignan, CNRS/IN2P3 -- Universit{\'e} Bordeaux I, F-33175 Gradignan Cedex, France}
\author{J. G. Johansen}
\affiliation{Department of Physics and Astronomy, Aarhus University, DK-8000 Aarhus C, Denmark}
\author{A. Jokinen}
\affiliation{Department of Physics, University of Jyv{\"a}skyl{\"a}, FIN-40351 Jyv{\"a}skyl{\"a}, Finland}
\author{B. Jonson}
\affiliation{Fundamental Fysik, Chalmers Tekniska H{\"o}gskola, S-41296 G{\"o}teborg, Sweden}
\author{T. \surname{Kurturkian-Nieto}}
\affiliation{Centre d'{\'E}tudes Nucl{\'e}aire de Bordeaux-Gradignan, CNRS/IN2P3 -- Universit{\'e} Bordeaux I, F-33175 Gradignan Cedex, France}
\author{J. H. Kusk}
\affiliation{Department of Physics and Astronomy, Aarhus University, DK-8000 Aarhus C, Denmark}
\author{T. Nilsson}
\affiliation{Fundamental Fysik, Chalmers Tekniska H{\"o}gskola, S-41296 G{\"o}teborg, Sweden}
\author{A. Perea} 
\affiliation{Instituto de Estructura de la Materia, CSIC, E-28006 Madrid, Spain}
\author{V. Pesudo}
\affiliation{Instituto de Estructura de la Materia, CSIC, E-28006 Madrid, Spain}
\author{E. Picado}
\affiliation{Grupo de F{\'i}sica Nuclear, Universidad Complutense, E-28040 Madrid, Spain}
\author{K. Riisager}
\affiliation{Department of Physics and Astronomy, Aarhus University, DK-8000 Aarhus C, Denmark}
\author{A. Saastamoinen}
\altaffiliation[Present address: ]{Cyclotron Institute, Texas A\&M University, College Station, TX 77843-3366, USA}
\affiliation{Department of Physics, University of Jyv{\"a}skyl{\"a}, FIN-40351 Jyv{\"a}skyl{\"a}, Finland}
\author{O. Tengblad}
\affiliation{Instituto de Estructura de la Materia, CSIC, E-28006 Madrid, Spain}
\author{J.-C. Thomas}
\affiliation{GANIL, CEA/DSM-CNRS/IN2P3, F-14076 Caen Cedex 5, France}
\author{J. \surname{Van de Walle}}
\affiliation{CERN, CH-1211 Geneva 23, Switzerland}

\date{\today}

\begin{abstract}
Resonances just above the proton threshold in \S affect the \rec reaction under astrophysical conditions. The $(p,\gamma)$-reaction rate is currently determined indirectly and depends on the properties of the relevant resonances.
We present here a method for finding the ratio between the proton and $\gamma$ partial widths of resonances in \S. The widths are determined from the $\beta 2p$- and $\beta p\gamma$-decay of \Ar, which is produced at the ISOLDE facility  at the European research organization CERN.
Experimental limits on the ratio between the proton and $\gamma$ partial widths for astrophysical relevant levels in \S have been found for the first time.
A level at $\SIE{4689.2}{24}{keV}$ is identified in the $\gamma$ spectrum, and an upper limit on the \GG ratio of $\num{0.26}$ ($\SI{95}{\%}$ C.L.) is found. In the two-proton spectrum two levels at $\SIE{5227}{3}{keV}$ and $\SIE{5847}{4}{keV}$ are identified. These levels were previously seen to $\gamma$ decay and upper limits on the \GGr ratio of $\num{0.5}$ and $\num{9}$, respectively, ($\SI{95}{\%}$ C.L.) are found, where the latter differs from previous calculations.
\end{abstract}



\maketitle

\section{Introduction}
\label{introduction}


Detailed knowledge of the energy levels of exotic nuclei, especially the ones just above the thresholds for particle emission, is important for understanding astrophysical processes such as explosive hydrogen burning. \S is situated close to the proton drip line and is produced in the \rec reaction in the r$p$- and $\alpha p$-process in type I x-ray bursts \cite{jose2010,Astro174}. The relatively long life time of \S makes it a critical waiting-point nucleus for these processes \cite{fisker2004}.

The \rec reaction is also interesting for the study of presolar dust grains. The most extensively studied grains are \ce{SiC} grains, because they are relatively abundant. A small fraction of these have been suggested to originate from classical novae \cite{Astro551}. They are characterized by low $\ce{^{12}C}/\ce{^{13}C}$ and $\ce{^{14}N}/\ce{^{15}N}$ ratios, high $\ce{^{30}Si}/\ce{^{28}Si}$ ratios, and $\ce{^{29}Si}/\ce{^{28}Si}$ ratios close to or lower than terrestrial values. The silicon isotopic abundance can provide information on the dominant nuclear synthesis paths followed by the thermonuclear runaway, which sets in near the base of the accreted layers from a main sequence star onto a white dwarf in a binary system \cite{Astro612}. In order to understand the origin of the isotopic ratios observed, the processes that create and destroy the different silicon isotopes have to be well understood. One of these is the \rec reaction. 
If this reaction is faster than the $\beta^+$ decay of \P, the amount of \ce{^{30}Si} would increase and the amount of \ce{^{29}Si} would decrease \cite{Astro142}. 

The Gamow window of the \rec reaction for temperatures relevant for astrophysics spans $\SI{100}{keV}$ to $\SI{1100}{keV}$. This, along with the proton separation energy of \S being $\SIE{4395.6}{7}{keV}$ \cite{mass}, implies that the levels in \S interesting for astrophysics lie below $\SI{6}{MeV}$.


Iliadis \textit{et al.} \cite{Astro134} predicted that the reaction rate of the \rec reaction was dominated by two resonances in \S with spins $3^+$ and $2^+$ and excitation energies of $\SIE{4733}{40}{keV}$ and $\SIE{4888}{40}{keV}$, respectively, which had not been observed at that time. The first experimental evidence came through studies of $\ce{^{32}S}(p,t)\S$ by Bardayan \textit{et al.} \cite{Bardayan} and Setoodehnia \textit{et al.} \cite{Seto82}. The most recent studies of \S can be found in Refs. \cite{Seto82, Seto83, Seto2012, Lotay2012, Almaraz2012}, while the results from previous experiments are combined in Ref. \cite{M30}.

The reaction rate is calculated using the energy, spin, and the proton and $\gamma$ partial width of the relevant resonances. The improvements on the energies of the $3^+$ and $2^+$ levels have reduced the uncertainties of the \rec reaction rate significantly for the relevant temperatures. The calculations made by Setoodehnia \textit{et al.} \cite{Seto2012} shows that the uncertainties are now so small that they no longer significantly influence the silicon abundances. These calculations, however, use proton and $\gamma$ partial widths, which are calculated based on the shell model and comparisons with the mirror nucleus. Experimental values for these would clearly be preferred. We present here a method for finding the ratio between the proton and $\gamma$ partial widths for the resonances of astrophysical interest using the decay of \Ar. Until now the preferred method for studying these resonances has been by the use of reaction experiments. The results presented here open up a new approach. Further results from our experiment will be published separately.

The experiment is described in \sref{theexperiment}. \sref{results} presents the results of the analysis of the low-lying states of \S and compares then with results in recent papers. It includes the method for finding the ratio between the proton and $\gamma$ partial widths. Finally, \sref{summary} summarizes the main results.

\section{The experiment}
\label{theexperiment}
\begin{figure}%
	\centering
	\includegraphics[width=0.90\columnwidth]{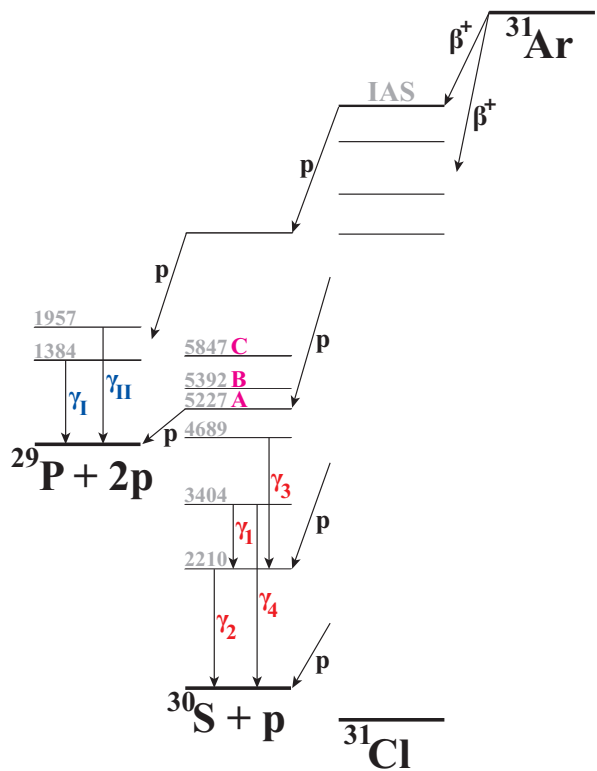} 
	\caption{(Color online) The $\beta$ decay of \Ar. Different proton decays are drawn as an illustration. The marked proton levels in \S correspond to \fref{s302p} and the marked $\gamma$ transitions correspond to \fref{gspec}. }%
	\label{level}%
\end{figure} 

In this experiment, the resonances in \S are studied via the $\beta$-delayed proton-$\gamma$ decay and the $\beta$-delayed two-proton decay of \Ar. The $\beta$-delayed two-proton decay is known to be mainly sequential \cite{Fynbo2p}. A partial decay scheme of \Ar is shown in \fref{level}. 
The experiment was optimized for detection of the delayed two-proton decay through a compact setup including detectors capable of stopping high-energy protons. This gave substantial $\beta$ background at low energy, which along with electronic noise made it challenging to identify low-energy protons.

\begin{figure}%
	\centering
	\includegraphics[width=0.7\columnwidth]{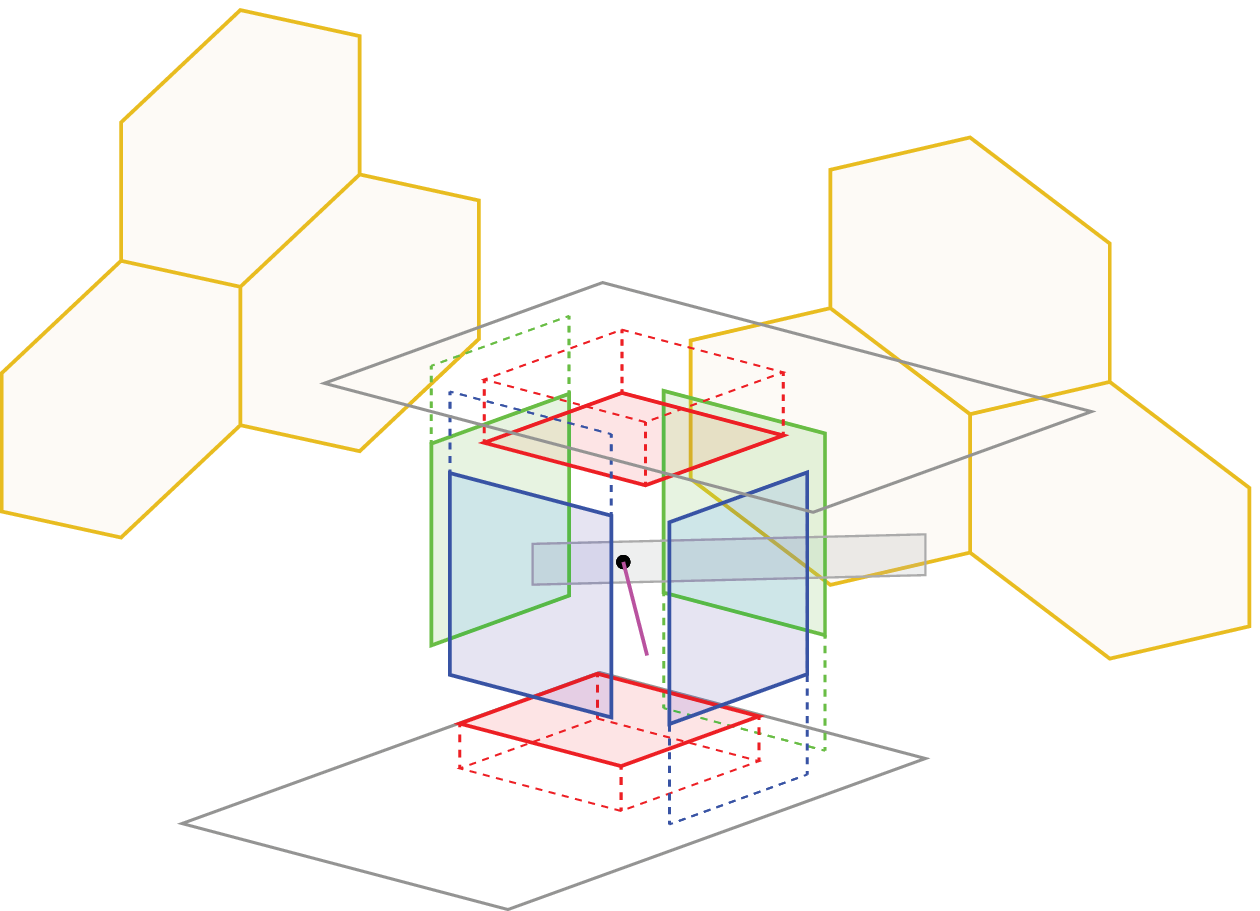} \\
	\vspace{0.5cm}
	\includegraphics[width=0.7\columnwidth]{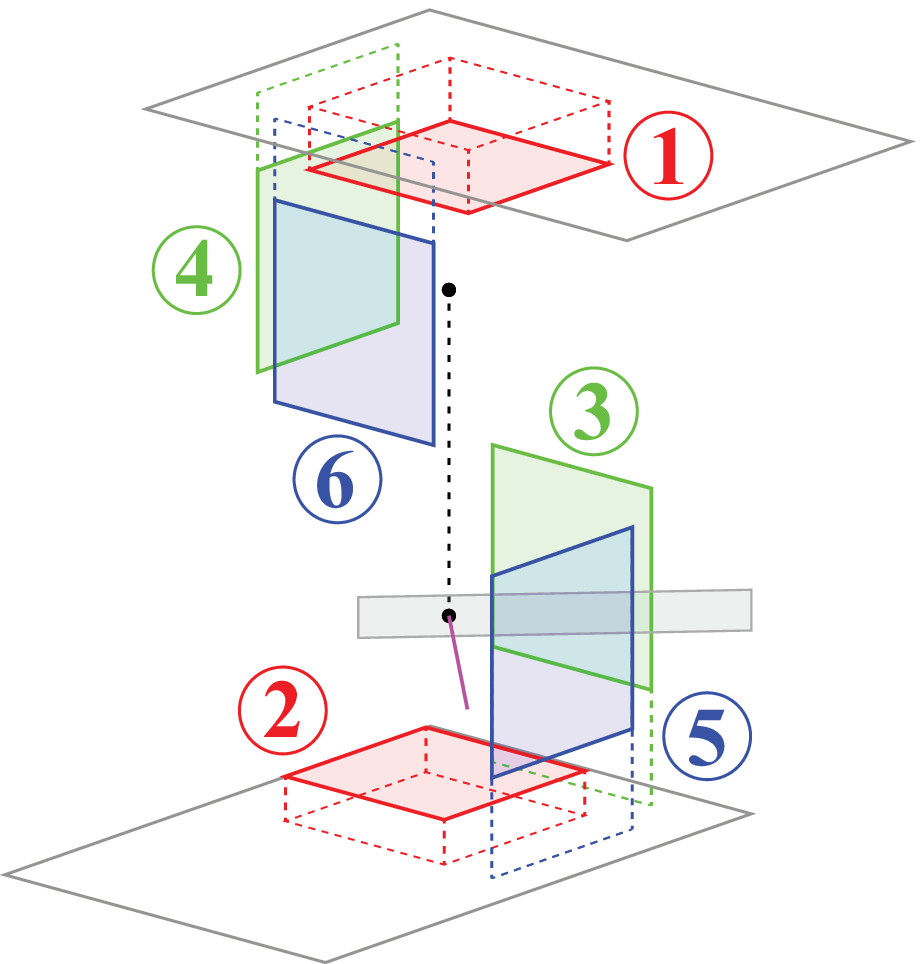}
	\caption{(Color online) The experimental setup used for the experiment. The beam enters between DSSSD 5 and 6 and is stopped in a foil mounted on a small metal holder entering between DSSSD 3 and 5. Two clustered germanium detectors situated outside the cube behind DSSSD 3 and 4 are shown on the upper drawing. The top of the cube with three of the DSSSDs is lifted, following the dotted black line, for better visualization on the bottom drawing.}%
	\label{setup}%
\end{figure} 

The radioactive $\SIb{60}{keV}$ \Ar beam used in the experiment was produced at the ISOLDE facility  at the European research organization CERN using a \ce{CaO} target and a versatile arc discharge plasma ion source \cite{CaOtarget} coupled to the General Purpose Separator (GPS) \cite{ISOLDE}. The beam is produced by irradiating the target with short high-intensity proton pulses at low repetition rate \cite{ISOLDE}. An average \Ar yield of about $\num{1}$ per second was obtained for a runtime of about $\num{7}$ days.

The $\SIb{60}{keV}$ beam was collected in a $\SIb{50}{\mu g/cm^2}$ carbon foil situated in the middle of the detector setup consisting of the silicon cube detector \cite{cube}, containing six double sided silicon strip detectors (DSSSDs): one $\SI{69}{\mu m}$ (1), one $\SI{494}{\mu m}$ (5), and four close to $\SI{300}{\mu m}$ (2--4, 6). The setup can be seen in \fref{setup}. The DSSSDs are segmented into $\num{16}$ strips in the front and in the back side, each $\SI{3}{mm}$ wide and $\SI{0.1}{mm}$ apart. Behind the thin DSSSD (1) and three of the four $\SIb{300}{\mu m}$ DSSSDs (2, 3, 6) $\SI{50}{mm}\times\SI{50}{mm}$ unsegmented silicon pad detectors (thickness close to $\SI{1500}{\mu m}$) were placed, which enables particle identification. In some part of the analysis only the four DSSSDs backed by a pad detector are used. For detection of $\gamma$ rays, two cluster detectors, each consisting of three germanium crystals, from MINIBALL \cite{miniball} were placed outside the chamber containing the silicon cube.

The energy and geometry calibration of the DSSSDs are made with \ce{^{33}Ar} produced from the same target-ion source unit as \Ar. A number of runs were made with \ce{^{33}Ar} during the experiment to monitor if the energies drifted. \ce{^{33}Ar} decays by $\beta$-delayed proton emission with well-known proton energies. For calibration of DSSSD 3--6 three known levels of \ce{^{33}Cl} were used: $\SIE{3971.24}{19}{keV}$, $\SIE{4112.34}{20}{keV}$, and $\SIE{5548.6}{5}{keV}$ \cite{ndsA33}, corresponding to proton energies of $\SIE{1642.8}{5}{keV}$, $\SIE{1779.5}{5}{keV}$, and $\SIE{3171.9}{7}{keV}$. DSSSD 2 has several broken strips and thus fewer counts. The protons from the two lowest \ce{^{33}Cl} resonances used could not be identified here. Instead, two higher-lying proton peaks are used. Their energies were found in the calibrated proton spectra of DSSSD 3 and 6, which had a better energy resolution, giving proton energies of $\SIE{2479.2}{20}{keV}$ and $\SIE{3856.9}{20}{keV}$. Due to the thickness of DSSSD 1 protons with energies above $\SI{2.50}{MeV}$ were not stopped inside the detector. For calibration of this detector the \ce{^{33}Cl} resonances at $\SIE{3971.24}{19}{keV}$ and $\SIE{4112.34}{20}{keV}$ were used together with two proton peaks found from DSSSD 3 and 6 at $\SIE{1320.1}{20}{keV}$ and $\SIE{2479.2}{20}{keV}$.

The solid angle of the DSSSDs is $\SI{43}{\%}$ of $4\pi$, or, if the two detectors without backing are disregarded, $\SI{27}{\%}$ of $4\pi$. Protons with energies below $\SI{500}{keV}$ can be stopped inside the collection foil depending on their emitted angle. The low-energy protons, which are interesting for this work, have an energy around $\SI{280}{keV}$. The angular coverage of these protons is $\SI{20}{\%}$ of $4\pi$, when the two detectors without backing are disregarded. 

The half-life of \Ar is just $\SIE{14.1}{7}{ms}$ \cite{Fynbo2p}. The beam gate for \Ar was therefore only open for $\SI{100}{ms}$ after a proton pulse has hit the target. In addition to this a software time window from $\SI{5}{ms}$ to $\SI{100}{ms}$ after proton impact was used. 

The pad detectors are calibrated with a \ce{^{148}Gd} source and a triple $\alpha$ source consisting of \ce{^{241}Am}, \ce{^{239}Pu}, and \ce{^{244}Cm}. 

The Ge detectors were calibrated using first a \ce{^{137}Cs} and a \ce{^{60}Co} source and then improved using a \ce{^{152}Eu} source, and lines from the decays of \ce{^{32,33}Ar} and \ce{^{16,18}N} were recorded online. The absolute $\gamma$ efficiency was found using a relative efficiency curve determined in a slightly different detector configuration (using four different $\gamma$ sources: \ce{^{152}Eu}, \ce{^{60}Co}, \ce{^{207}Bi} and \ce{^{11}Be}) and an absolute measurement with an \ce{^{152}Eu} source. The result, using the formula in Ref. \cite{miniballeff}, is
\begin{align}
	\varepsilon_{\gamma} \left( E\right)  =& 0.21 \exp\Bigg(-2.669 - 1.457\log\left( \frac{E}{\SI{}{MeV}}\right) \nonumber\\ 
	 &- 0.231 \left[\log\left( \frac{E}{\SI{}{MeV}}\right) \right]^2  \Bigg) ,
\end{align}
with an estimated uncertainty of $\SI{10}{\%}$.

\section{Results and discussion}
\label{results}

\begin{table*}[tbp]
\caption{Energy and spin of the \S levels below $\SI{6}{MeV}$ for the present work and recent previous work. The proton separation energy of \S is $\SIE{4395.6}{7}{keV}$ \cite{mass}.}
\begin{tabular}{c c c c c c c c c}
\toprule[0.1em]
\vspace{-3mm} \\
\multicolumn{3}{c}{Setoodehnia \textit{et al}. \cite{Seto2012}}			 	& \multicolumn{2}{c}{Lotay \textit{et al}. \cite{Lotay2012}}		& \multicolumn{3}{c}{Almaraz-Calderon \textit{et al}. \cite{Almaraz2012}}		& Present work		 \\
 \cmidrule[0.05em](rl{1.5mm}){1-3} \cmidrule[0.05em](rl{1.5mm}){4-5} \cmidrule[0.05em](rl{1.5mm}){6-8} \cmidrule[0.05em](rl{1.5mm}){9-9}
 \vspace{-3mm} \\
 \vspace{0.8mm}
$\ce{^{28}Si}(\ce{^{3}He},n\gamma)\ce{^{30}S}$ & $\ce{^{32}S}(p,t)\ce{^{30}S}$ &		 	& \multicolumn{2}{c}{$\ce{^{28}Si}(\ce{^{3}He},n\gamma)\ce{^{30}S}$}		& $\ce{^{32}S}(p,t)\ce{^{30}S}$ & $\ce{^{28}Si}(\ce{^{3}He},n)\ce{^{30}S}$ &		& 	$\ce{^{31}Ar}(\beta^+)(p)\ce{^{30}S}$\\
$E_x$ (keV)	& $E_x$ (keV)	& $J^{\pi}$	& $E_x$ (keV) 	& $J^{\pi}$ 	&$E_x$ (keV)	& $E_x$ (keV) 	& $J^{\pi}$ 	& $E_x$ (keV) 	\\
\midrule[0.08em]
\vspace{-3mm} \\
g.s.				&	    	 		&	$0^+$ 		&	g.s.				&	$0^+$ 		& g.s.				&	g.s.	 		& 			 		&	g.s.		\\
2210.6(3)		& 	2208(3)		&$2^+$		& 2210.1(1)		& $2^+$ 		& 2208.5(22)   	&	2200(210)	&  				&	2210.2(11)	\\
3403.6(6)		& 			 		&$2^+$		& 3404.1(1)		& $2^+$ 		& 3405.8(12)		& 	   				& 				&	3404.4(16)	\\
3667.0(5)		& 			 		&$0^+$		& 3668.0(4)		& $0^+$ 		& \multirow{2}{*}{\centering 3677.3(70) }	&	3600(260)	 		&		&				\\
3676.9(4)		& 	3681(3)		&$1^+$		& 3677.1(4)		& $1^+$ 		& 						&			 		&  				&				\\
4688.1(4)		& 	4688(2)		&$3^+$		& 4687.6(2)		& $3^+$ 		& 4682.5(57)		& 			 		&			 		&	4689.2(24)	\\
4809.8(5)		& 	4812(2) 	&$2^+$		& 4808.7(3)		& $2^+$ 		& 						& 			 		&			 		&				\\
5132.3(5)		& 			 		&$(4^+)$		& 5132.1(1)		& $4^+$ 		& 5130.0(18)		&			 		& $(4^+)$ 		&				\\
			 		&5225(2)			& $(0^+)$		& 5218.8(3)		& $3^+$ 		& 5217.8(28)		&	5200(44)	& $(0^+)$ 		&	5227(3) \\
			 		&5315(2)			& $(3^-)$		& 	{}					& $(3^-)$		& 5312.1(20)		&			 		& $(3^-)$ 		&				 \\
			 		&5393(2)			& $3^+$		& 		{}				& $(2^+)$		& 5382.0(7)		&	5400(43)	& $(2^+)$ 		&	5392(4)\\ 
			 		&5849(2)			& $(2^+)$		&	5848.0(4)		& $4^+$ 		& 5835.5(13)		&			 		& $(4^+)$ 		&	5847(4) \\
 \vspace{0.4mm}			 		&5947(3)			& $(4^+)$		& 	{}					& 		{}		 	& 						& 			 		&			 		&				\\	
\bottomrule[0.1em]
\end{tabular}
\label{s30tab}
\end{table*}

\begin{figure}[btp]%
	\centering
	\includegraphics[width=0.99\columnwidth]{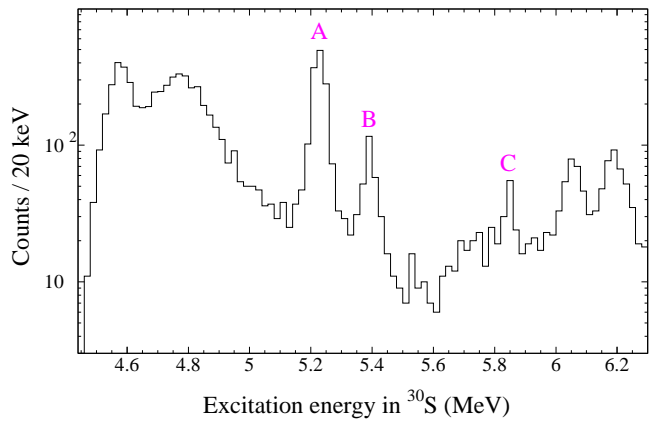} 
	\caption{(Color online) The excitation energy of \S calculated from the energies of the two protons from the sequential two-proton decay as described in Ref. \cite{Fynbo2p}. The energies of the marked peaks can be found in \tref{s30tab}.}%
	\label{s302p}%
\end{figure}

\begin{figure}[btp]%
	\centering
	\includegraphics[width=0.99\columnwidth]{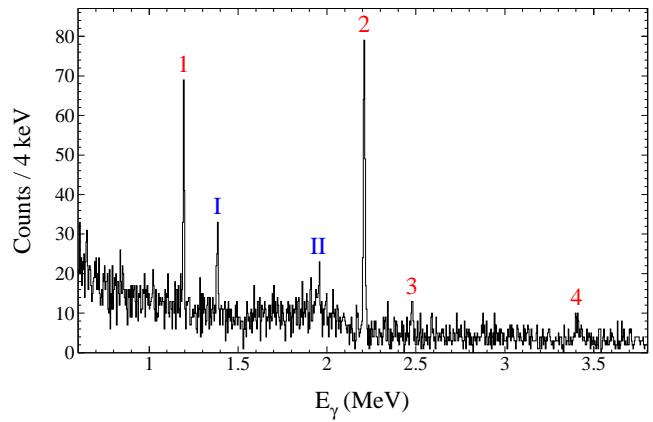} 
	\caption{(Color online) The summed $\gamma$ spectrum over all crystals gated on protons from the decay of \Ar. The numbers correspond to transitions in \S and the Roman numbers correspond to transitions in \P. The energies of the marked peaks and their relative intensities can be found in \tref{gtab}.}%
	\label{gspec}%
\end{figure} 

\begin{table}[bthp]
\caption{The energies and relative intensities of the $\gamma$ lines in \fref{gspec} normalized to the transition from the first excited state to the ground state of \S.}
\begin{tabular}{c c c}
\toprule[0.1em]
\vspace{-3mm} \\
 \vspace{0.8mm}Peak 		& Energy ($\SI{}{keV}$)		 &	Intensity		\\
\midrule[0.08em]
\vspace{-3.5mm} \\
1			&  $1194.2(11)$	& $35(4)$ 		\\
2			& $2210.1(11)$	 		& $100(8)$ 		\\
3			& $2478.9(21)$	 		& $16(4)$ 		\\
4			& $3407(7)$			& $22(7)$ 	 \\
I			& $1383.7(14)$	 	& $19(4)$ 		 \\
 \vspace{0.4mm} II			& $1957(4)$			& $10(4)$ 		\\
\bottomrule[0.1em]
\end{tabular}
\label{gtab}
\end{table}

\tref{s30tab} summarizes the latest published results on the low-lying levels of \S. Included are also the energies measured in this experiment. \fref{s302p} shows the levels above $\SI{5}{MeV}$ found, in the present work, using the energies of the two protons from the sequential two-proton decay as described in Ref. \cite{Fynbo2p}. The two levels above $\SI{6}{MeV}$, seen in \fref{s302p}, corresponds to previously identified levels \cite{Fynbo2p} and will be treated in an other publication. Due to contamination from $\beta$ particles and electronic noise, the levels below $\SI{5}{MeV}$ cannot be identified in the two-proton decay. These levels are known to decay mainly by $\gamma$ emission and are thus identified from the proton-gated $\gamma$ spectrum presented in \fref{gspec}. The relative intensities of the $\gamma$ lines are given in \tref{gtab}. Only the $\num{1194.2}$- and the $\SIb{2210.1}{keV}$ lines have previously been observed in the decay of \Ar \cite{Axelsson98b}. The $\num{1194.2}$- and the $\SIb{3407}{keV}$ lines both correspond to decays of the second excited state in \S. Their relative intensity is $\num{0.63}(\num{21})$, 
which is consistent within $2\sigma$ with the value of $\num{0.25}(\num{4})$ measured in Ref. \cite{Kuhlmann}. However, the difference of these values might imply a double structure of peak 4 in our spectrum. In the following we focus on the relative $\gamma$ and proton decay widths of the $\num{4689.2}$-, $\num{5227}$- and $\SIb{5847}{keV}$ levels. An overview of previously calculated \GGr ratios can be found in \tref{gamgam}.

\subsection*{The $\SIb{4689}{keV}$ level}

\begin{table*}[th]
\caption{Ratios between calculated proton and $\gamma$ partial widths from previous work compared with the $\SI{95}{\%}$ confidence limits extracted from the present work.}
\begin{tabular}{l l l l l l l l}
\toprule[0.1em]
\vspace{-3mm} \\
\multicolumn{3}{c}{Setoodehnia \textit{et al}. \cite{Seto82}}			 	& \multicolumn{3}{c}{Almaraz-Calderon \textit{et al}. \cite{Almaraz2012}}		& \multicolumn{2}{c}{Present work	}	 \\
 \cmidrule[0.05em](rl{1.5mm}){1-3} \cmidrule[0.05em](rl{1.5mm}){4-6} \cmidrule[0.05em](rl{1.5mm}){7-8}
 \vspace{-3mm} \\
$E_x$ (keV)	&$J^{\pi}$	& 		\GGr	 				& $E_x$ (keV) 	&$J^{\pi}$	&  		\GGr			 & $E_x$ (keV) 	&  \GGr		\\
\midrule[0.08em]
\vspace{-3mm} \\
4699(6)			&$3^+$ 	&	 $\num{210}$ 				& 				& 			 	&									& 4689.2(24) & 	$>3.8$	\\
4814(3)			&$2^+$ 	&	 $\num{1.3}$ 				& 				& 			 	&									& 				& 				\\
5136(2)			&$3^+$ 	&	 $\num{480}$ 				& 5130		&$4^+$ 	&	 $\num{340}$ 			& 				& 				\\
5217.8(14) 	&$0^+$ 	& $\geq\num{0.36e-3}$	& 5217.8	& $0^+$	& $\num{0.7014e-3}$	& 5227(3)	& 	$<0.5$	\\
5314(7)			&$3^-$ 	&	$\num{9.8e-3}$ 			& 5312.1	&$3^-$ 	&	 $\num{4.6}$ 			& 		 		& 				\\	
5391(3)			&$2^+$	& $\num{0.26e-2}$ 			& 5382		&$2^+$ 	&$\num{3.8954e-2}$	& 				& 				\\
					&				&	 							 		& 5835.5	&$4^+$ 	&	 $\num{15.7}$ 			& 5847(4) 	& 	$<9$	\\
\bottomrule[0.1em]
\end{tabular}
\label{gamgam}
\end{table*}

The level at $\SIE{4689.2}{24}{keV}$ is observed for the first time in the decay of \Ar. It is identified in the $\gamma$ spectrum (\fref{gspec}) as peak 3, which corresponds to the decay of the level to the first excited state. It is predicted to decay by proton emission as well, but due to contamination from $\beta$ particles and electronic noise it is not possible to identify it in \fref{s302p}. $\beta$ particles mainly deposit a small amount of energy in the DSSSD and give a larger signal in the pad detector.
Since this source of background cannot be identified in the DSSSDs without backing, they are omitted in the following  analysis.
 
The large background below $\SI{5}{MeV}$ in \fref{s302p} consists of multiplicity two events where one of the signals is caused by a $\beta$ or electronic noise typically in coincidence with a proton from a strong one-proton peak, e.g., a proton from \Cl to the ground state in \S (see \fref{level}). The background can thus be reduced by including only multiplicity two events with a proton that is known to feed the level considered. The protons feeding the $\SIb{4689}{keV}$ level are found by gating on the $\SIb{2478.9}{keV}$ $\gamma$. 
The protons are found to have energies in the intervals $[\num{1580},\num{1920}]\SI{}{keV}$ and $[\num{2200},\num{2400}]\SI{}{keV}$. By gating on protons with these energies the background is reduced substantially. The $\SIb{4689}{keV}$ level, however, is still not positively identified in the two-proton spectrum and thus only an upper limit on the \GG ratio can be found. The background can be estimated by gating on protons in the interval $[\num{2040},\num{2120}]\SI{}{keV}$; they correspond to a known strong proton group at $\SIE{2084}{2}{keV}$ from \Cl to the ground state of \S \cite{Fynbo2p}. The background spectrum then has to be scaled by a factor $f=\num{0.311}(\num{4})$, namely to the ratio between the number of protons gated on in the interesting intervals compared to the number of protons gated on for the background.
The number of two-proton events, giving a \S energy in a $\SIb{40}{keV}$ interval around $\SI{4689}{keV}$, is found to be $(\num{26}-f\num{33})$. 
Using the same proton gates, the number of $\gamma$ rays, corresponding to the decay of the $\SIb{4689}{keV}$ level to the first excited state in \S, is found to be $(\num{13}-f\num{11})$.

Due to the limited number of counts in the spectra it is necessary to determine the upper limit through a simulation.
The upper limit is found from an ensemble, which is made by drawing four numbers $n_i$ from a probability function $P(n_i|n_{0i})$, where $n_{0i}$ are the four numbers found from the spectra ($n_{01}=\num{26}$, $n_{02}=\num{33}$, $n_{03}=\num{13}$, $n_{04}=\num{11}$). This probability function is derived as follows: $n_{0i}$ is a number from a Poisson distribution with mean value $\lambda_i$ and the $n_i$'s needed to create the ensemble are random numbers drawn from a Poisson distribution with this mean value $\lambda_i$. The problem arises because $\lambda_i$ is unknown. Therefore, one has to integrate over all the possible values of $\lambda_i$ weighted with the probability of this $\lambda_i$ given $n_{0i}$:
\begin{align}
	P(n_i|n_{0i}) &= \frac{\int_{0}^{\infty}d\lambda_i P(n_i|\lambda_i)P(\lambda_i|n_{0i})}{\int_0^{\infty}d\lambda_i P(\lambda_i|n_{0i})} \nonumber\\
	&= \int_0^{\infty}d\lambda_i P(n_i|\lambda_i)P(n_{0i}|\lambda_i) \\
	&=\frac{(n_i+n_{0i})!}{n_i!n_{0i}!}\frac{1}{2^{n_i+n_{0i}+1}} , \nonumber
\end{align}
where the second equality comes from Bayes's theorem that employing a uniform prior distribution gives $P(\lambda_i|n_i)=P(n_i|\lambda_i)$ and from the Poisson distribution being normalized in $\lambda_i$. 

The \GG ratio can then be found from these four random numbers $n_i$ drawn from the four distributions by correcting for efficiencies:
\begin{align}
	\frac{\Gamma_p}{\Gamma_{\gamma}} = \frac{(n_1-f n_2)/\epsilon_p}{(n_3-f n_4)/\epsilon_{\gamma}},
\end{align}
where the $\gamma$ efficiency $\epsilon_{\gamma}$ includes the $\gamma$ intensities measured by Lotay \textit{et al}. \cite{Lotay2012}. From the ensemble of $10^6$ \GG values found in this way, we find a $\SI{95}{\%}$ confidence upper limit of $\num{0.26}$ on the \GG ratio. 

This limit can be compared to calculations made by Setoodehnia \textit{et al}. \cite{Seto82} for a $3^+$ resonance at $\SI{4699}{keV}$. They find: ${\Gamma_p}/{\Gamma_{\gamma}}=\num{0.47e-2}$. Recent developments with nano structured CaO targets have increased the yield of \Ar up to an order of magnitude \cite{Joao}. Hence, using this type of target and a setup optimized for low-energy protons, it should be possible to identify the proton decay of the $\SIb{4689}{keV}$ level using the gating technique presented here and thereby deduce an experimental value for the \GG ratio.

\subsection*{The $\SIb{5227}{keV}$ level}
A level at $\SIE{5218.8}{3}{keV}$ has been observed to $\gamma$ decay to the first excited state of \S by a $\SIEb{3008.5}{2}{keV}$ $\gamma$ and to the second excited state by a $\SIEb{1814.4}{3}{keV}$ $\gamma$ with relative branching ratios of $\num{0.80}(\num{9})$ and $\num{0.20}(\num{12})$, respectively \cite{Lotay2012}. These lines are not seen in the $\gamma$ spectrum. Assuming this is the same level as we have identified in the two-proton spectrum at $\SIE{5227}{3}{keV}$, an upper limit on the \GGr ratio can be found by gating on the protons feeding the level. These protons are found by gating on the $\SIb{5227}{keV}$ peak in the two-proton spectrum and choosing the proton with the highest energy as this is most likely to be the first emitted proton.   
The two detectors without backing are again excluded. The protons feeding the level are found to have energies in the intervals $[\num{1000},\num{1350}]$ and $[\num{1670},\num{1950}]\SI{}{keV}$. The background is estimated by gating on another strong transition to the ground state of \S with proton energies in the interval $[\num{1380},\num{1460}]\SI{}{keV}$ (the proton group at $\SIE{1416}{2}{keV}$ in Ref. \cite{Fynbo2p}). The number of two-proton events in the \S peak at $\SI{5227}{keV}$ is $\num{226}\pm\num{15}$. 
The are no $\gamma$ rays in a $\SIb{50}{keV}$ interval around $\SI{3008}{keV}$, corresponding to the $\gamma$ decay to the first excited state in \S. This gives a $\SI{95}{\%}$ upper value for the number of $\gamma$ rays of $\num{2.7}$ \cite{limit}. Including the $\gamma$ intensities from Ref. \cite{Lotay2012}, this gives a $\SI{95}{\%}$ confidence upper limit of the \GGr ratio of $\num{0.5}$.

As seen in \tref{gamgam} Setoodehnia \textit{et al}. \cite{Seto82} and Almaraz-Calderon \textit{et al}. \cite{Almaraz2012} estimated for a $0^+$ level at $\SI{5217.8}{keV}$ that ${\Gamma_{\gamma}}/{\Gamma_p}\geq\num{0.36e-3}$ and ${\Gamma_{\gamma}}/{\Gamma_p}=\num{0.7014e-3}$, respectively. Almaraz-Calderon \textit{et al}. \cite{Almaraz2012} found a proton branching ratio of $\num{1.00}(2)$ for a level at $\SIE{5200}{44}{keV}$, while Lotay \textit{et al}. \cite{Lotay2012} observed $\gamma$ decays of a level at $\SIE{5218.8}{3}{keV}$. These results are hard to reconcile if the same state was populated. It should be noted that the calculations by Setoodehnia \textit{et al}. and Almaraz-Calderon \textit{et al}. are made for a $0^+$ state, but it is not clear whether the level observed around $\SI{5220}{keV}$ is $0^+$ or $3^+$, or if there are in fact two levels in this energy region.

\subsection*{The $\SIb{5847}{keV}$ level}
The level at $\SIE{5847}{4}{keV}$ has been observed to $\gamma$ decay to the first excited state of \S by a $\SIEb{3637.7}{4}{keV}$ $\gamma$ ray \cite{Lotay2012}. There is no clear evidence of such a line in the $\gamma$ spectrum and an upper limit on the \GGr ratio can thus be found as described for the $\SIb{5227}{keV}$ level. 
The two detectors without backing are again excluded. The protons feeding the level are found to have energies in the intervals $[\num{2220},\num{3400}]$ and $[\num{5540},\num{6160}]\SI{}{keV}$. The background is chosen as for the $\SIb{4689}{keV}$ level. The number of two-proton events in the \S peak at $\SI{5847}{keV}$ is $\num{24}\pm\num{5}$. 
The number of $\gamma$ rays in a $\SIb{50}{keV}$ interval around $\SI{3638}{keV}$ is $(\num{2}-f\num{12})$, where $f=\num{0.463}(4)$ defined as above. This gives a $\SI{95}{\%}$ upper value for the number of $\gamma$ rays of $\num{2.88}$ \cite{limit}. After correcting for efficiencies this gives a $\SI{95}{\%}$ confidence upper limit of the \GGr ratio of $\num{9}$.

Almaraz-Calderon \textit{et al}. \cite{Almaraz2012} do not observe any significant proton branch from this level in their $\ce{^{28}Si}(\ce{^{3}He},n)\ce{^{30}S}$ experiment. This does not agree with our result of a proton branching ratio of at least $\num{0.1}$ ($\SI{95}{\%}$ C.L.). They assume it to be a $4^+$ state and estimate that: ${\Gamma_{\gamma}}/{\Gamma_p}=\num{15.7}$, in contradiction to our upper value of $\num{9}$ ($\SI{95}{\%}$ C.L.).


\section{Summary}
\label{summary}
The levels below $\SI{6}{MeV}$ in \S have been studied by the $\beta 2p$- and  $\beta p\gamma$-decay of \Ar.

The $\gamma$ decay of the second excited state of \S to the ground state has been observed for the first time in the decay of \Ar. The relative intensities of the $\gamma$ lines corresponding to this decay and the decay to the first excited state have been found to be $\num{0.63}(\num{21})$.

The $\gamma$ decay of the astrophysically interesting level at $\SIE{4689.2}{24}{keV}$ has also been observed for the first time in the decay of \Ar.

We present a new analysis method that provides experimental limits on the ratio between the proton and $\gamma$ partial widths of resonances in \S. The upper limit of the \GG ratio has been found for the level at $\SIE{4689.2}{24}{keV}$ to be $\num{0.26}$ ($\SI{95}{\%}$ C.L.), and upper limits of the \GGr ratio of $\num{0.5}$ and $\num{9}$ ($\SI{95}{\%}$ C.L.) have been found for levels at $\SIE{5227}{3}{keV}$ and $\SIE{5847}{4}{keV}$, respectively. The latter conflicts with previous calculations.



\section{Acknowledgement}
This work was supported by the European Union Seventh Framework through ENSAR (Contract No. 262010). This work was partly supported by the Spanish Funding Agency under Projects No. FPA2009-07387, No. FPA2010-17142, and No. AIC-D-2011-0684, by the French ANR (Contract No. ANR-06-BLAN-0320), and by R{\'e}gion Aquitaine. A.S. acknowledges support from the Jenny and Antti Wihuri Foundation.

\flushleft 
\bibliography{ar31}

\end{document}